\documentclass[ reprint, amsmath,amssymb, aps,pra]{revtex4-1}
\usepackage{tikz}
\usepackage{amscd}
\usepackage{amsthm}
\usepackage{graphicx}
\usepackage{mathdots}
\usepackage{epstopdf}
\usepackage{enumerate}
\usepackage{changepage}

\usepackage[
colorlinks,
linkcolor = blue,
citecolor = blue,
urlcolor = blue]{hyperref}
\def \qed {\hfill \vrule height7pt width 7pt depth 0pt}

\newtheorem{theorem}{Theorem}
\newtheorem{corollary}{Corollary}
\newtheorem{lemma}{Lemma}
\newtheorem{example}{Example}
\newtheorem{definition}{Definition}
\usepackage{dcolumn}
\usepackage{bm}

\begin{document}

\preprint{APS/123-QED}

\title{Masking quantum information in multipartite scenario}

\author{Mao-Sheng Li}
 \affiliation{
 Department of Mathematical Sciences, Tsinghua University, Beijing100084, China}
\author{Yan-Ling Wang}%
 \email{wangylmath@yahoo.com}
\affiliation{
 School of Computer Science and Network Security, Dongguan University of Technology, Dongguan 523808, China
}

\begin{abstract}
Recently, Kavan Modi \emph{et al.} found that masking quantum information is impossible in bipartite scenario. This adds another item of the  no-go theorems.  In this paper, we present some new schemes different from error correction codes, which show that quantum states can be masked when more participants are allowed in the masking process. Moreover, using a pair of mutually orthogonal Latin squares of dimension $d$, we show that all the $d$ level quantum states can be masked into tripartite quantum systems whose local dimensions are $d$ or $d+1$.   This highlight  some difference between the no-masking theorem and the classical no-cloning  theorem or no-deleting theorem.

	\begin{description}
\item[PACS numbers] 03.67.Hk,03.65.Ud
\end{description}
 \end{abstract}                            
\maketitle

\section{Introduction}
 It is fundamental  to find out the boundary between the classical and quantum information. Entanglement plays an  important role in quantum information\cite{Horodecki09}.  Due to the existence of  entanglement, the quantum information exhibit some surprising power such as quantum teleportation \cite{Bennett93,Bouwmeester97} and quantum key distribution\cite{Bennett92,Gisin02}.  The quantum process is modeled by some unitary operator. As a consequence,  unlike our classical information, no quantum machine can perfectly copy an unknown quantum state.  This is one of the well known no-go theorems: no-cloning theorem\cite{James70,Wootter82,Dieks82}.  Another example shows   the difference between the classical and quantum information is the Bell inequality  \cite{EPR,nils}.

 Classical information can be hidden in quantum correlation of a bipartite state. Recently,  Kavan Modi \emph{et al.} \cite{Kavan18} asked   whether quantum information can be stored  only in the quantum correlations between two quantum systems rather than the system itself.  An coding  process is called masking, which can make the original information inaccessible to both local systems.  Based on the   linearity and unitarity  of the quantum mechanics, they highlighted another no-go theorem: no-masking theorem. That is,  it is impossible to mask an arbitrary states into bipartite quantum systems.  As a by-product, this result could deduce the impossibility of the quantum qubit commitment, which is a stronger version of the well known quantum bit commitment\cite{Mayers97,Lo97}. In addition, they also showed that a set of states that can be masked is helpful for quantum secret sharing \cite{Hillery99,Cleve99}  and   other potential applications in quantum communication protocols in the future.

However, there are still many questions to be answered about the problem of masking. For examples, how to determine the maskable states of a given physical masker? If we hide the original quantum information in the mixed states rather than the pure ones, what results can we obtain?
In Ref. \cite{Kavan18}, the authors pointed out that it is possible to mask quantum information into multipartite quantum states by quantum error correction codes.  Are there any other interesting hiding methods besides quantum error correcting codes?

Here we show some new schemes related to this subject.  In Sec. \ref{second}, we give some necessary definitions about the masking of quantum information and some  related  concepts. In  Sec. \ref{simplecase}, we study how to mask quantum information in multipartite scenario. Firstly,  we illustrate how to mask all the qubit (qutrit) states  in a four-qubit (six-qutrit) system by  two simple examples.    Following the spirit of the two constructions, it is easy to generalize to arbitrary higher quantum system.  In fact, we show that for any positive integer $d\geq 2$, we can mask all the quantum states in $\mathbb{C}^d$ by adding $2d-1$ systems of the same dimension in  the masking scenario.   In  Sec. \ref{tripartitecase},  we study the  optimal scheme on numbers of  parties in order to mask arbitrary quantum states.   We show that tripartite quantum systems is enough to  achieve this task.  In fact, using a pair of mutually orthogonal Latin squares of dimension $d$, we show that all the $d$ level quantum states can be masked into tripartite quantum systems with local dimension  of $d$. Finally, we draw some conclusions and put forward some interesting questions in the last section.

\section{Preliminaries }\label{second}
In this section, we  mainly give some necessary definitions about the masking of quantum information and some  related  concepts which will be used later.
Firstly, instead of hiding quantum information in bipartite quantum system, we hide them in multipartite ones. So we need to   generalize the definition of masking of quantum states which is almost the same as  that in \cite{Kavan18}.

\begin{definition}\label{masking} An operation $\mathcal{S}$ is said to  mask quantum information contained in states $\{|a_k\rangle_{A_1}\in\mathcal{H}_{A_1}\}$ by mapping them to states $\{|\Psi_k\rangle\in \bigotimes_{j=1}^n \mathcal{H}_{A_j}\}$ such that all the marginal states  of $|\Psi_k\rangle$ are identical; i.e.,
$$\rho_{A_j}=\text{Tr}_{\widehat{A_j}}(|\Psi_k\rangle \langle \Psi_k|),\ j\in\{1,2,...,n\}$$
where $\widehat{A_j}$ denotes the set $\{A_1,A_2,...,A_n\}\setminus \{A_j\}.$
\end{definition}
In fact, the masker should be modeled by a unitary operator $U_\mathcal{S}$ on $A_1$ plus some ancillary systems $\{A_2,...,A_n\}$ and  given by
$$ \mathcal{S}:\ \  U_\mathcal{S}|a_k\rangle_{A_1}\otimes|\beta\rangle_{\widehat{A_1}}=|\Psi_k\rangle.$$

The  masker $\mathcal{S}$ on $\mathcal{H}_{A_1}$ is completely determined by the effect the unitary operator $U_\mathcal{S}$ acting on a base of $\mathcal{H}_{A_1}$. Let $|0\rangle, |1\rangle,..., |d-1\rangle$ be an orthonormal basis of  $\mathcal{H}_{A_1}$. Suppose $U_\mathcal{S}|j\rangle_{A_1}\otimes|\beta\rangle_{\widehat{A_1}}=|\Phi_j\rangle$ for $0\leq j\leq d-1$.  Through the article, we use the following  simplify notation to denote the above masking process.
$$|j\rangle\rightarrow|\Phi_j\rangle, \ \ j\in\{0,1,...,d-1\}.$$

It is mentioned in \cite{Kavan18} that it is possible  to mask an arbitrary quantum state with more than two parties are allowed.
 There they pointed out that some error correction code \cite{Shor95,Lidar13} represented such an example.    In fact, the famous Shor's 9-qubits code \cite{Shor95}:
$$
  \begin{array}{c}
    |0\rangle\rightarrow \frac{|000\rangle+|111\rangle}{\sqrt{2}}\otimes\frac{|000\rangle+|111\rangle}{\sqrt{2}}\otimes\frac{|000\rangle+|111\rangle}{\sqrt{2}}, \\
  |1\rangle\rightarrow \frac{|000\rangle-|111\rangle}{\sqrt{2}}\otimes\frac{|000\rangle-|111\rangle}{\sqrt{2}}\otimes\frac{|000\rangle-|111\rangle}{\sqrt{2}}. \\
  \end{array}
 $$
 If we denote the first state to be $|\Psi_0\rangle$ and $|\Psi_1\rangle$  for the second one, then the general qubit state $\alpha_0|0\rangle+\alpha_1|1\rangle$ should be changed into $\alpha_0|\Psi_0\rangle+\alpha_1|\Psi_1\rangle$. It is straightforward to check that all the local states of the multiqubit state  are equal to  $I_2/2$.  Hence, we   deduce that all the qubit states can be masked by 9 qubit systems as the above process.

In combinatorics and in experimental design, a Latin square of dimension $d$  is an $d\times d$ array filled with $d$ different symbols (for instance $\{1,2,...,d\}$) such that each symbol appears   in each row and  in each column precisely  once.  We also call such a matrix   as  a Latin square with order $d$.  An example of a $3\times 3$ Latin square is
$$
\left[
  \begin{array}{ccc}
    1 & 2 & 3 \\
    3 & 1 & 2 \\
    2 & 3 & 1 \\
  \end{array}
\right].
$$
\begin{definition}  Two Latin squares $V=(V_{jk}),W=(W_{jk})$  of dimension $d$  are called    orthogonal, if the following equation holds
  $$ \{(V_{jk},W_{jk})\ \ \big| \ \ 1\leq j, k\leq d\}=\{(j,k) \ \big|\ 1\leq j,k\leq d\}.$$
\end{definition}

A family of pairwise orthogonal Latin squares is normally called mutually orthogonal
Latin squares, and abbreviated ``MOLS ". The maximum size of a family of MOLS of
order $d$ is denoted $N(d)$.


\section{masking quantum states into multipartite  quantum systems}\label{simplecase}
To move forward, we first give another simple example to mask all the qubit states without using the error correcting code.

\begin{example}\label{qubit} All the qubit states can be masked by the processing defined by
$$
  \begin{array}{c}
    |0\rangle\rightarrow |\Psi_0\rangle=\frac{|00\rangle+|11\rangle}{\sqrt{2}}\otimes\frac{|00\rangle+|11\rangle}{\sqrt{2}}, \\
    |1\rangle\rightarrow |\Psi_1\rangle=\frac{|00\rangle-|11\rangle}{\sqrt{2}}\otimes\frac{|00\rangle-|11\rangle}{\sqrt{2}}. \\
  \end{array}
 $$
 \end{example}

\noindent\emph{Proof:} Let $\vec{\alpha}=(\alpha_0,\alpha_1)$, the general qubit state $\alpha_0|0\rangle+\alpha_1|1\rangle$ should be changed into $|\Psi_{\vec{\alpha}}\rangle=\alpha_0|\Psi_0\rangle+\alpha_1|\Psi_1\rangle$ under the above process. In fact, $|\Psi_{\vec{\alpha}}\rangle$ can be written as the following simple form:
$$\frac{(\alpha_0+\alpha_1)}{2}(|0000\rangle+|1111\rangle)+\frac{(\alpha_0-\alpha_1)}{2}(|0011\rangle+|1100\rangle).
$$
Since $|000\rangle,|111\rangle,011\rangle, |100\rangle $ are orthonormal, we can apply  lemma \ref{partial_trace} in APPENDIX A to deduce the  partial trace of $|\Psi_{\vec{\alpha}} \rangle \langle \Psi_{\vec{\alpha}}|$   as follows
$$
  \begin{array}{ccl}
    \rho_{A_1}& = & \text{Tr}_{\widehat{A_1}}(|\Psi_{\vec{\alpha}}\rangle \langle \Psi_{\vec{\alpha}}|) \\[2mm]
      &= & \frac{1}{4} (|\alpha_0+\alpha_1|^2+|\alpha_0-\alpha_1|^2)(|0\rangle\langle 0|+|1\rangle\langle 1|) \\[2mm]
      &=&\frac{1}{2}(|0\rangle\langle 0|+|1\rangle\langle 1|)=\frac{I_2}{2}.
  \end{array}
$$
where we use the fact that $|\alpha_0|^2+|\alpha_1|^2=1$.  By the symmetry of the four systems, $\rho_{A_j}=I_2/2$ for all $j$.  \qed

\begin{example}\label{qutrit} Define $|\Psi_0\rangle,|\Psi_1\rangle,|\Psi_2\rangle$ to be
{$$
  \begin{array}{c}
\frac{|00\rangle+|11\rangle+|22\rangle}{\sqrt{3}}\otimes\frac{|00\rangle+|11\rangle+|22\rangle}{\sqrt{3}}\otimes\frac{|00\rangle+|11\rangle
+|22\rangle}{\sqrt{3}}, \\[2mm]
  \frac{|00\rangle+\omega|11\rangle+\omega^2|22\rangle}{\sqrt{3}}\otimes \frac{|00\rangle+\omega|11\rangle+\omega^2|22\rangle}{\sqrt{3}}\otimes \frac{|00\rangle+\omega|11\rangle+\omega^2|22\rangle}{\sqrt{3}}, \\[2mm]
    \frac{|00\rangle+\omega^2|11\rangle+\omega|22\rangle}{\sqrt{3}}\otimes \frac{|00\rangle+\omega^2|11\rangle+\omega|22\rangle}{\sqrt{3}}\otimes \frac{|00\rangle+\omega^2|11\rangle+\omega|22\rangle}{\sqrt{3}}
  \end{array}
 $$}
\noindent respectively, here $\omega=e^{\frac{2\pi i}{3}}.$
 All the qutrit states can be masked by the processing defined by
 $$|0\rangle\rightarrow|\Psi_0\rangle,\ \  |1\rangle\rightarrow|\Psi_1\rangle,\ \ |2\rangle\rightarrow|\Psi_2\rangle.$$
\end{example}

\noindent\emph{Proof:} Let $\vec{\alpha}=(\alpha_0,\alpha_1, \alpha_2)$ be a unit vector. The general qubit state $\alpha_0|0\rangle+\alpha_1|1\rangle+\alpha_2|2\rangle$ should be changed into $|\Psi_{\vec{\alpha}}\rangle=\alpha_0|\Psi_0\rangle+\alpha_1|\Psi_1\rangle+\alpha_2|\Psi_2\rangle$ under the above process. In order to calculate the local states, we need an explicit expression of the  global state $|\Psi_{\vec{\alpha}}\rangle$.  We notice that there are 27 terms in the expansion of $|\Psi_{\vec{\alpha}}\rangle$.  Each term is of the form $|j_0j_0\rangle|j_1j_1\rangle|j_2j_2\rangle$ with $0\leq j_0,j_1,j_2\leq 2.$ In order to determine the coefficient corresponding to $|j_0j_0\rangle|j_1j_1\rangle|j_2j_2\rangle$  in $|\Psi_{\vec{\alpha}}\rangle$, we   find out the corresponding contributions of $\alpha_0|\Psi_0\rangle,\alpha_1|\Psi_1\rangle,\alpha_2|\Psi_2\rangle $ are just
$$\frac{1}{3}\frac{\alpha_0}{\sqrt{3}}, \ \ \ \ \frac{1}{3}\frac{\omega^{j_0+j_1+j_2}\alpha_1}{\sqrt{3}},\ \ \ \frac{1}{3}\frac{\omega^{2(j_0+j_1+j_2)}\alpha_2}{\sqrt{3}}.$$

We find that there are only three kinds of coefficients in the expansion of $|\Psi_{\vec{\alpha}}\rangle$.   In table \ref{exam1}, we list the terms by column  whose coefficients corresponding to the first element of the same column.
\begin{table}[h]

$\begin{array}{ccc}\hline\hline

 \frac{ 1 }{3} \frac{\alpha_0+\alpha_1+\alpha_2}{\sqrt{3}}&\frac{ 1 }{3} \frac{\alpha_0+\omega\alpha_1+\omega^2\alpha_2}{\sqrt{3}} & \frac{ 1 }{3} \frac{\alpha_0+\omega^2\alpha_1+\omega\alpha_2}{\sqrt{3}} \\ \hline
 |00\rangle|00\rangle|00\rangle& |00\rangle|00\rangle|11\rangle&|00\rangle|00\rangle|22\rangle\\
 |00\rangle|11\rangle|22\rangle& |00\rangle|11\rangle|00\rangle&|00\rangle|11\rangle|11\rangle\\
 |00\rangle|22\rangle|11\rangle& |00\rangle|22\rangle|22\rangle&|00\rangle|22\rangle|00\rangle\\
 |11\rangle|00\rangle|22\rangle& |11\rangle|00\rangle|00\rangle&|11\rangle|00\rangle|22\rangle\\
 |11\rangle|11\rangle|11\rangle& |11\rangle|11\rangle|22\rangle&|11\rangle|11\rangle|00\rangle\\
 |11\rangle|22\rangle|00\rangle& |11\rangle|22\rangle|11\rangle&|11\rangle|22\rangle|22\rangle\\
 |22\rangle|00\rangle|11\rangle& |22\rangle|00\rangle|22\rangle&|22\rangle|00\rangle|00\rangle\\
 |22\rangle|11\rangle|00\rangle& |22\rangle|11\rangle|11\rangle&|22\rangle|11\rangle|22\rangle\\
 |22\rangle|22\rangle|22\rangle& |22\rangle|22\rangle|00\rangle&|22\rangle|22\rangle|11\rangle\\
  \hline\hline

\end{array}$
\caption{This table shows the coefficients of the expansion of $|\Psi_{{\vec{\alpha}}}\rangle$ . For example, the coefficient corresponding to the term
$|11\rangle|22\rangle|11\rangle$ is $\frac{ 1 }{3} \frac{\alpha_0+\omega\alpha_1+\omega^2\alpha_2}{\sqrt{3}}$.}\label{exam1}

\end{table}

 Since the above 27 terms are orthonormal without consider the first system $A_1$,    applying lemma \ref{partial_trace} in APPENDIX A  we can easily calculate the partial trace of $|\Psi_{\vec{\alpha}} \rangle \langle \Psi_{\vec{\alpha}}|$
{\small$$
  \begin{array}{ccl}
    \rho_{A_1}& = & \text{Tr}_{\widehat{A_1}}(|\Psi_{\vec{\alpha}}\rangle \langle \Psi_{\vec{\alpha}}|) \\[2mm]
      &= & \frac{3}{9} (\displaystyle\sum_{j=0}^{2}\big| \frac{\sum_{k=0}^{2}\omega^{jk}\alpha_k}{\sqrt{3}}\big|^2)(|0\rangle\langle 0|+|1\rangle\langle 1|+|2\rangle\langle 2|)\\[2mm]
      &=&\frac{1}{3}I_3.
  \end{array}
$$  }
The third equality can be obtained by observing $|\alpha_0|^2+|\alpha_1|^2+|\alpha_2|^2=1$  and  the unitarity of
$$\frac{1}{\sqrt{3}}\left[
  \begin{array}{ccc}
    1 & 1 & 1 \\
    1 & \omega & \omega^2 \\
    1 & \omega^2 & \omega \\
  \end{array}
\right]$$
which preserves the length of vector $(\alpha_0, \alpha_1, \alpha_2)^t$.
That is, we have {\footnotesize$$|\frac{\alpha_0+\alpha_1+\alpha_2}{\sqrt{3}}|^2+|\frac{\alpha_0+\omega\alpha_1+\omega^2\alpha_2}{\sqrt{3}}|^2+ | \frac{\alpha_0+\omega^2\alpha_1+\omega\alpha_2}{\sqrt{3}}|^2=1.$$
}

By the symmetry of the six systems, $\rho_{A_j}=I_3/3$ for all $j$. Therefore, regardless of how the three parameters of $\alpha_0, \alpha_1, \text{ and } \alpha_2 $ are selected,  all six local states share the same information about $|\Psi_{\vec{\alpha}}\rangle$.   This is just the definition of the masker we defined before. \qed

\vskip 3pt

Note that the  masking process  above is different from the process derived from   error correcting  code  in \cite{Cleve99}.  Moreover, from the proof above, we can do much more.
Now we are in the situation to write down our main theorem:

\begin{theorem}\label{general} For any positive integer $d\geq 2$, we can mask all the quantum states in $\mathbb{C}^d$ by adding $2d-1$ systems of the same dimension. That is, they can be masked in $\bigotimes_{j=1}^d \mathcal{H}_j$ with all $\mathcal{H}_j=\mathbb{C}^d.$
\end{theorem}
\noindent\emph{Proof}: Set $\omega=e^{\frac{2\pi i}{d}} $ and let $|0\rangle, |1\rangle,...,|d-1\rangle$ be an orthogonal  normalized  basis of  $\mathbb{C}^d$. Now we define the unitary processing as
$$
|l\rangle\rightarrow|\Psi_l\rangle=\bigotimes_{j=1}^d\frac{\sum_{k=0}^{d-1} \omega^{kl}|kk\rangle}{\sqrt{d}}, \ l\in\{0, 1,...,d-1\}.
$$
The general  state $|\vec{\alpha}\rangle=\sum_{l=0}^{d-1}\alpha_l|l\rangle$ should be changed into $|\Psi_{\vec{\alpha}}\rangle=\sum_{l=0}^{d-1}\alpha_l|\Psi_l\rangle$ under the above process.
 We notice that there are $d^d$ terms in the expansion of $|\Psi_{\vec{\alpha}}\rangle$.  Each term is of the form $|j_0j_0\rangle|j_1j_1\rangle\cdots|j_{d-1}j_{d-1}\rangle$ with $0\leq j_0,j_1,\cdots, j_{d-1}\leq d-1.$ In order to determine the coefficient corresponding to $|j_0j_0\rangle|j_1j_1\rangle\cdots|j_{d-1}j_{d-1}\rangle$  in $|\Psi_{\vec{\alpha}}\rangle$, if setting  $j=\sum_{l=0}^{d-1}j_l$,  we calculate the corresponding contribution of $\alpha_0|\Psi_0\rangle,\alpha_1|\Psi_1\rangle,\cdots, \alpha_{d-1}|\Psi_{d-1}\rangle $ are
{ $$\frac{1}{d^{\frac{d-1}{2}}}\frac{\alpha_0}{\sqrt{d}}, \ \ \ \ \ \frac{1}{d^{\frac{d-1}{2}}}\frac{\omega^{j}\alpha_1}{\sqrt{d}},\ \ \cdots, \ \ \frac{1}{d^{\frac{d-1}{2}}}\frac{\omega^{j(d-1) }\alpha_{d-1}}{\sqrt{d}}$$
}
\noindent respectively. Hence the coefficient of $|\Psi_{\vec{\alpha}}\rangle$  corresponding to the term  $|j_0j_0\rangle|j_1j_1\rangle\cdots|j_{d-1}j_{d-1}\rangle$  is just $$ \frac{1}{d^{\frac{d-1}{2}}}\frac{\sum_{k=0}^{d-1}\omega^{jk}\alpha_k}{\sqrt{d}}.$$

Noticing that $\omega$ is a primitive $d$-th root of unit.  So $\omega^d=1$. And  hence $\omega^{\sum_{l=0}^{d-1}j_l}$ is  just determined by $\sum_{l=0}^{d-1}j_l \text{ mod } d$. Hence fixing any $j_0,j_1,\cdots, j_{d-2}$, we have set equality
$$\{\sum_{l=0}^{d-1}j_l \text{ mod } d \ \big | 0 \leq j_{d-1}\leq d-1\}=\{0, 1, ..., d-1\}.$$
Moreover, if we  fix $j_0\in \{0,1\cdots, d-1\}$, there are $d^{d-2}$ choices for the $j_1,\cdots, j_{d-2}$. With the two noticing points and using lemma \ref{partial_trace} in APPENDIX A, we can obtain the following partial trace
$$
{
  \begin{array}{ccl}
    \rho_{A_1}& = & \text{Tr}_{\widehat{A_1}}(|\Psi_{\vec{\alpha}}\rangle \langle \Psi_{\vec{\alpha}}|) \\[2mm]
      &= & \frac{d^{d-2}}{d^{d-1}} (\displaystyle\sum_{j=0}^{d-1}\big| \frac{\sum_{k=0}^{d-1}\omega^{jk}\alpha_k}{\sqrt{d}}\big|^2)(\sum_{l=0}^{d-1}|l\rangle\langle l|) \\[2mm]
      &=&\frac{1}{d}I_d.
  \end{array}
  }
$$
The third equality can be obtained by observing $|\alpha_0|^2+|\alpha_1|^2+\cdots+|\alpha_{d-1}|^2=1$  and  the unitarity of
$$\frac{1}{\sqrt{d}}\left[
                 \begin{array}{ccccc}
                   1 & 1 & 1 & \cdots & 1 \\
                   1 & \omega  & \omega^2 & \cdots & \omega^{d-1} \\
                   1 & \omega^2 & \omega^4 & \cdots  & \omega^{2(d-1)} \\
                   \vdots & \vdots & \vdots & \ddots & \vdots \\
                   1 & \omega^{d-1} & \omega^{2(d-1)} & \cdots & \omega^{(d-1)^2}\\
                 \end{array}
               \right]
 $$
 which preserves the length of vector $(\alpha_0, \alpha_1, \cdots, \alpha_{d-1})^t$. Hence we have the following equality
$$ \displaystyle\sum_{j=0}^{d-1}\big| \frac{\sum_{k=0}^{d-1}\omega^{jk}\alpha_k}{\sqrt{d}}\big|^2=1.$$

The same as example \ref{qutrit}, by the symmetry of the $2d$ systems, $\rho_{A_j}=I_d/d$ for all $j$.
Therefore, regardless of how the $d$  parameters of $\alpha_0, \alpha_1, \cdots, \alpha_{d-1}$ are selected,  all the $2d$ local states share the same information about $|\Psi_{\vec{\alpha}}\rangle$.  This completes the proof.
\qed

\vskip 5pt

\noindent \textbf{Remark 1:} One of the referees points out   an alternative method for proving theorem \ref{general} which is much more concise. Now we borrow  it and present it here  as follows. Denote $d$  generalized Bell states $|\psi_k\rangle=\frac{1}{\sqrt{d}}\sum_{j=0}^{d-1} \omega^{jk}|jj\rangle$ for $k\in\{0,1,...,d-1\}$. Then $|\Psi_{\vec{\alpha}}\rangle=\sum_{k=0}^{d-1}\alpha_k|\psi_k\rangle^{\otimes d} $. Now using the orthogonal relations among the above $d$ generalized Bell states, i.e.  $\langle\psi_k|\psi_l\rangle= \delta_{kl}$, the  partial trace of $|\Psi_{\vec{\alpha}}\rangle$ over all but the first two parties leads to $\rho_{A_1A_2}:=\sum_{k=0}^{d-1}|\alpha_k|^2 |\psi_k\rangle\langle\psi_k|.$  Then using the property of maximally entangled states, i.e. $\text{Tr}_{A_2}(|\psi_k\rangle\langle \psi_k|)=I_d/d$ and the identity $\sum_{k=0}^{d-1}|\alpha_k|^2=1$, we can deduce $\text{Tr}_{A_2}(\rho_{A_1A_2})=I_d/d.$ To conclude, we have
$$\begin{array}{rcl}
  \text{Tr}_{\widehat{A_1}}(|\Psi_{\vec{\alpha}}\rangle\langle\Psi_{\vec{\alpha}}|)
 & = & \text{Tr}_{A_2}(\text{Tr}_{\widehat{A_1A_2}}(|\Psi_{\vec{\alpha}}\rangle\langle\Psi_{\vec{\alpha}}|))\\[1mm]
 & = & \text{Tr}_{A_2}(\rho_{A_1A_2})=I_d/d.
\end{array}
$$
Therefore, using the above argument and the symmetry property of given states, we can also arrive at the same conclusion. \qed

\vskip 2pt

The states we constructed above share some similar property of the bipartite maximally entangled states: all the local states are completely mixed. We notice that these states are multipartite entangled. Hence multipartite entangled states are useful for quantum states masking process. However, multipartite entanglement is  still hard to characterize  in quantum information theory\cite{Huber17,Vicente13}.

\section{masking quantum states into tripartite quantum systems}\label{tripartitecase}
In this section, we show that the optimal parties in order to mask an arbitrary  quantum states is tripartite.
Suppose $\{|1\rangle, |2\rangle, \cdots, |d\rangle\}$ is an orthonormal basis of $\mathbb{C}^d$.

 Sets of Latin squares that are orthogonal to each other have found   application in the classical  error correcting codes.  Instead of dealing with the classical information, here we show that a pair of orthogonal  Latin squares are also useful for    masking  quantum information.

 In order to denote our target  masking process in tripartite systems, we only need three matrices. Given two mutually orthogonal Latin squares $V=(v_{jk}),W=(w_{jk})$ whose dimension  is $d$, we  show how to use this two matrices to construct a masking process. Firstly, we define $U$ to be another $d\times d$ matrix whose $jk$-th entry is just $k$. Noticing that $U$ is not a Latin square!  Then we can define a physical process $\mathcal{U}_{V,W}$ as follows
\begin{equation}\label{processing}
\begin{array}{rcl}
  \mathbb{C}^d & \rightarrow & \mathbb{C}^d\otimes\mathbb{C}^d\otimes\mathbb{C}^d\\
  |j\rangle & \mapsto &|\Phi_j\rangle=  \frac{1}{\sqrt{d}}\sum_{k=1}^d|u_{jk}v_{jk}w_{jk}\rangle,  \ \ j=1, 2,...,d.
\end{array}
\end{equation}

For example, if $d=4$ and $U, V,W$ are chosen to be the following matrices respectively

$$\left[
  \begin{array}{cccc}
    1 & 2 & 3 & 4 \\
    1 & 2 & 3 & 4 \\
    1 & 2 & 3 & 4 \\
     1 & 2 & 3 & 4
  \end{array}
\right],\ \ \
\left[
  \begin{array}{cccc}
    1 & 2 & 3 & 4 \\
    2 & 1 & 4 & 3 \\
    3 & 4 & 1 & 2\\
   4 &3 & 2 & 1 \\
  \end{array}
\right],\ \ \
\left[
  \begin{array}{cccc}
    1 & 2 & 3 & 4 \\
    4 & 3 & 2 & 1 \\
    2 & 1 & 4 & 3\\
    3 & 4 & 1 & 2 \\
  \end{array}
\right].$$
Then the  encoding process defined above is just
{\small$$
  \begin{array}{crl}
   \alpha|1\rangle+\beta|2\rangle+\gamma|3\rangle+\delta|4\rangle & \mapsto &   \frac{\alpha}{2}(|111\rangle+|222\rangle+|333\rangle+|444\rangle) \\[1mm]
      & + & \frac{\beta}{2}( |124\rangle+|213\rangle+|342\rangle+|431\rangle)\\[1mm]
      & +  & \frac{\gamma}{2}( |132\rangle+|241\rangle+|314\rangle+|423\rangle)\\[1mm]
      & + & \frac{\delta}{2}( |143\rangle+|234\rangle+|321\rangle+|412\rangle).
  \end{array}
$$
}
  \begin{theorem}\label{masking_latin_square} Let $d$ be an integer greater than 2, that is, $d\geq 3 $ and $d\in \mathbb{N}$. If there exist $V,W\in M_{d}(\mathbb{C})$ such that they  are orthogonal Latin squares labeling by symbols  $\{1,2,...,d\}$. Then    all the $d$ level quantum states can be masked by the process defined in (\ref{processing}) into
tripartite systems $\mathbb{C}^d\bigotimes\mathbb{C}^d\bigotimes\mathbb{C}^d.$
 \end{theorem}
\noindent \emph{Proof}: Define $U$   to  be a $d\times d$ matrix whose $jk$-th entry is just $j$. Since $V,W\in M_{d}(\mathbb{C})$ are orthogonal Latin squares,  it is easy to  verify the following set equalities
\begin{enumerate}[(a)]
  \item $
    \{(u_{jk},v_{jk})\ \ \big| \ \ 1\leq j, k\leq d\}=\{(j,k) \ \big|\ 1\leq j, k\leq d\},$
 $$\{w_{mk} \ \ \big| \ \  1\leq  k\leq d\}=\{ k  \ \big|\ 1\leq  k\leq d\};$$

  \item
$
    \{(u_{jk},w_{jk})\ \ \big| \ \ 1\leq j, k\leq d\}=\{(j,k) \ \big|\ 1\leq j, k\leq d\},$
 $$\{v_{mk} \ \ \big| \ \  1\leq  k\leq d\}=\{ k  \ \big|\ 1\leq  k\leq d\};$$

  \item
$
    \{(v_{jk},w_{jk})\ \ \big| \ \ 1\leq j,k\leq d\}=\{(j,k) \ \big|\ 1\leq j,k\leq d\},$
 $$\{u_{mk} \ \ \big| \ \  1\leq  k\leq d\}=\{ k  \ \big|\ 1\leq  k\leq d\};$$
\end{enumerate}
for each $m\in\{1,2,...,d\}$. By the definition of the process  $\mathcal{U}_{V,W}$ in (\ref{processing}), $\mathcal{U}_{V,W}$ transfers $\sum_{j=1}^d\alpha_j|j\rangle$  to
$$|\Phi\rangle:=\sum_{j=1}^d\alpha_j|\Phi_j\rangle= \sum_{j=1}^d\sum_{k=1}^d\frac{\alpha_j}{\sqrt{d}}|u_{jk}v_{jk}w_{jk}\rangle .$$ Noticing the  equalities in (\textcolor[rgb]{0.00,0.00,1.00}{a}) and then applying  lemma \ref{partial_trace} in APPENDIX A, we have
{\small$$\text{Tr}_{AB}(|\Phi\rangle\langle\Phi| )=\sum_{j=1}^d \sum_{k=1}^d( \frac{|\alpha_j|^2}{d} |w_{jk}\rangle\langle w_{jk}|)=\sum_{j=1}^d ( \frac{|\alpha_j|^2}{d} I_d)=\frac{I_d}{d}.$$}
With similar argument and using equalities in (\textcolor[rgb]{0.05,0.29,0.93}{b}) and (\textcolor[rgb]{0.05,0.29,0.93}{c}), we deduce the other two partial traces
$$\text{Tr}_{AC}(|\Phi\rangle\langle\Phi| )=\text{Tr}_{BC}(|\Phi\rangle\langle\Phi| )=\frac{I_d}{d}.$$
Therefore, we can conclude that the defined process $\mathcal{U}_{V,W}$ is indeed a masking processing.\qed

\vskip 5pt

If $d$  is an odd integer greater than 2, we can easily present   a pair of mutually orthogonal Latin squares.  Define  $V,W$ to be the following  two matrices respectively:
$$V=(v_{jk})=\left[
  \begin{array}{cccccc}
    1 & 2 & 3 & \cdots & d-1 & d \\
    d & 1 & 2 & \cdots& d-2&d-1 \\
    d-1& d & 1 & \cdots &d-3&d-2 \\
     \vdots &  \vdots & \ddots &\ddots & \ddots &\vdots \\
     3& 4& 5&\cdots & 1&2\\
       2&   3& 4& \cdots & d&1
  \end{array}
\right]
$$
and
$$W=(w_{jk})=\left[
  \begin{array}{cccccc}
    1 & 2 & 3 & \cdots & d-1 & d \\
    2 & 3 & 4 & \cdots& d&1 \\
    3& 4& 5 & \cdots &1&2 \\
     \vdots &  \vdots & \ddots &\ddots & \ddots &\vdots \\
     d-1& d& 1&\cdots & d-3&d-2\\
       d&  1& 2& \cdots & d-2&d-1
  \end{array}
\right].
$$

By definition of $V,W$, we have  (here and below,  $n \text{ mod } d $ is taken from the equivalence class $\{1,2,...,d\}$ of module $d$)   $$(v_{jk},w_{jk})=(k-j+1 \text{ mod } d, j+k-1 \text{ mod } d ).$$
Fixed $v_{jk}=l$ with $l\in \{1,2,...,d\}$, for any $k$ chosen from $ \{1,2,...,d\}$,  there are exactly one $j\in \{1,2,...,d\}$ with condition  $k-j\equiv l-1 \text{ mod } d$.  At this point, $w_{jk}=j+k-1=2k-l \text{ mod } d$.
Noticing that if $d$ is odd, then for any integer $1\leq l\leq d$
\begin{equation}\label{setequal}
\{2k-l\text{ mod } d  \big|  \ k=1, 2,...,d\}=\{1,2,3,...,d\}
\end{equation}
With the   equality (\ref{setequal}), it can easy to deduce that
$$
\{(v_{jk},w_{jk})\ \ \big| \ \ 1\leq j,k\leq d\}=\{(l,s) \ \big|\ 1\leq l ,s\leq d\}.
$$
Therefore, $V$ and $W$ are indeed orthogonal Latin squares.

\begin{corollary}\label{masking_odd} If $d$ is an odd integer and $d\geq 3$, then all the $d$ level states can be masked in
tripartite systems $\mathbb{C}^d\bigotimes\mathbb{C}^d\bigotimes\mathbb{C}^d.$ As a consequence, if $d$ is even and $d\geq 2$, then all the $d$ level states can be masked in
tripartite systems $\mathbb{C}^{d+1}\bigotimes\mathbb{C}^{d+1}\bigotimes\mathbb{C}^{d+1}.$
 \end{corollary}

  \vskip 5pt
In 1779, Euler started looking at the problem of finding orthogonal Latin squares of every dimension. He conjectured that no Latin squares of dimension $d \equiv 2 \ (\text{mod } 4)$ exists. The first result casting serious doubts on the truth of Euler's conjecture is
due to Bose and Shrikhande (1959) who were able to construct  a pair of orthogonal Latin squares
dimension $d= 22$ \cite{Bose59}. And the final result due to Bose, Shrikhande and Parker,
proving the falsity of Euler's conjecture \cite{Dey13,Bose60} shows as follows:  There exists at least   a pair of orthogonal Latin squares of dimension $d$ when  $d>2$ and $d\neq 6$. Therefore, we have the following corollary.

\begin{corollary}\label{masking_odd} Suppose $d$ is an integer greater than $2$ and $d \neq 6$, then all the $d$ level states can be masked in
tripartite systems $\mathbb{C}^d\bigotimes\mathbb{C}^d\bigotimes\mathbb{C}^d.$
 \end{corollary}

\vskip 6pt

\noindent{\textbf{Remark 2:}} One could find that there is quantum version of Latin square which is known as quantum Latin squares\cite{Musto16,Musto17}. A quantum Latin square of order $d$ is an $d\times d$ array of elements of the Hilbert
space $\mathbb{C}^d$, such that every row and every column is an orthonormal basis. And quantum Latin squares have been found  application to construction of  unitary error bases and mutually unbiased basis. Using a pair of orthogonal quantum Latin squares one  can similarly  derive a masking process   with Theorem \ref{masking_latin_square}.  Since quantum Latin squares is a general form of Latin square, this provides more ways to gain  a masking process.

\vskip 4pt

In this section, we just study the optimal number of parties which is sufficient for masking an arbitrary quantum states of given quantum system. It is also interesting to find out the optimality of local dimensions when fixing  the number of parties. However, this is out of our ability so far. Hence we just leave it as questions in the conclusion for future study.

\section{conclusion and discussion}

In this paper, we study the processing for quantum information masking. Unlike the no masking theorem in bipartite system, it can mask   arbitrary quantum states when more participants taken part in the masking process.  Although this conclusion can be derived from the previously known results in \cite{Cleve99}, here we present some new schemes to show how to mask quantum information in multipartite scenario.    More precisely, we show that all the quantum states in $\mathbb{C}^d$  can be masked into an processing when more $2d-1$ systems of the same level  taken part in. It is interesting  to find out whether the number of systems or the level of system can be smaller in order to masking the same quantum information. Therefore,  we study the  optimal scheme on numbers of  parties in order to mask arbitrary quantum states.   We show that tripartite quantum systems is enough to  achieve this task. In fact, we show any pair orthogonal Latin squares of dimension $d$  are useful for masking $d$ level quantum states.

Noticing that all the known masking schemes, the local states of each partite is just  equal to $I_d/d$. Therefore, it is interesting to wonder whether it is possible to masking all the quantum states of $\mathbb{C}^d$ into tripartite quantum systems $\mathbb{C}^d\otimes\mathbb{C}^d\otimes\mathbb{C}^d$ such that   the marginal states are $\{\rho_A,\rho_B,\rho_C\}$  do not equal to $I_d/d$. Moreover, can all quantum states of level $d$  be hidden into tripartite quantum system $\mathbb{C}^n\otimes\mathbb{C}^n\otimes\mathbb{C}^n$ with $n<d$ or not?

\vskip 5pt

	\vspace{2.5ex}

	\noindent{\bf Acknowledgments}\, \, The authors are very grateful to the reviewer   for  providing us many useful and insightful suggestions.
	This work is supported by the  NSFC 11571119.
	
\vskip 10pt

	{
		$${\text{\textbf{APPENDIX A}}}$$
	}

\begin{lemma}\label{partial_trace} Let $ |\Psi\rangle$ be a quantum states in multipartite system  $\bigotimes_{j=1}^n \mathcal{H}_{A_j}$.  For any $j\in\{1,...,n\}$,
if $ |\Psi\rangle$ can be written as
$$ \sum_{k=1}^{n_j}c_k|\psi_k\rangle_{A_j}|\mu_k\rangle_{\widehat{A_j}}$$
with $\{|\mu_k\rangle_{\widehat{A_j}}: 1\leq k\leq n_j\}$ to be orthonormal states in the system without $A_j$. Then we have the following partial trace
 $$  \rho_{A_j} = \text{Tr}_{\widehat{A_j}}(|\Psi\rangle \langle \Psi|)=\sum_{k=1}^{n_j}|c_k|^2|\psi_k\rangle_{A_j}\langle\psi_k|.$$
 Here we use the same notation $\widehat{A_j}$ which   has been used in the definition \ref{masking}.
\end{lemma}
\noindent\emph{Proof:}  In fact, the following calculation is straight forward by the definition of the partial trace. The significant point here is the orthonormal property of the set $\{|\mu_k\rangle_{\widehat{A_j}}: 1\leq k\leq n_j\}$. So we have $\text{Tr}(|\mu_k\rangle_{\widehat{A_j}}\langle\mu_l|)=\delta_{kl}.$
 $$\begin{array}{ccl}
  \rho_{A_j} & = &\text{Tr}_{\widehat{A_j}}(|\Psi\rangle \langle \Psi|)   \\
    & = & \text{Tr}_{\widehat{A_j}}(\displaystyle\sum_{k=1}^{n_j}\sum_{l=1}^{n_j}c_k\overline{c_l}|\psi_k\rangle_{A_j}\langle\psi_l|\otimes|\mu_k\rangle_{\widehat{A_j}}\langle\mu_l| )  \\

     & =&\displaystyle\sum_{k=1}^{n_j}\sum_{l=1}^{n_j}c_k\overline{c_l}|\psi_k\rangle_{A_j}\langle\psi_l|\cdot \text{Tr}(|\mu_k\rangle_{\widehat{A_j}}\langle\mu_l|)  \\
         & =&\displaystyle\sum_{k=1}^{n_j}\sum_{l=1}^{n_j}c_k\overline{c_l}|\psi_k\rangle_{A_j}\langle\psi_l|\cdot\delta_{kl} \\
             & =&\displaystyle\sum_{k=1}^{n_j} |c_k|^2 |\psi_k\rangle_{A_j}\langle\psi_k|.
 \end{array}
 $$ \qed

\end{document}